\documentclass[proof,hidelinks]{WileyASNA-v2}
\articletype{Article Type}%
\received{00}
\revised{00}
\accepted{00}

\usepackage{amsmath}
\usepackage{placeins}
\usepackage{subcaption}
\usepackage{afterpage}

\newcommand{\mycomment}[1]{}

\begin{document}

\title{A Physical Model for the Ice Coma of the Interstellar, Hyperactive Comet 3I/ATLAS}

\author{Eric Keto}
\author{Abraham Loeb}
\authormark{Eric Keto}
\titlemark{Comet 3I/ATLAS}
\address{\orgdiv{Department of Astronomy}, \orgname{Harvard University},
\orgaddress{\state{Cambridge, MA}, \country{USA}}}
\corres{Eric Keto}
 \presentaddress{Center for Astrophysics, 60 Garden St, Cambridge, MA 02138}

 
\abstract[abstract]{ 
A previous study suggests that the observed exponential character of the surface brightness profiles in the coma around the interstellar comet 3I/ATLAS at 4 au can be explained as a consequence of the destruction of the icy scattering particles by sublimation. 
Here we follow the evolution of the ice coma as a function of heliocentric distance. 
We describe the evolution of the space and time distribution of the albedo within the coma 
by a Haser model for the fading of the grain albedo from a higher, icy value to a lower, refractory value.
The competing effects of increasing rates of production and sublimation produce a peak in the total scattering cross-section due to ice at a heliocentric distance of 3 - 4 au.
The modeled apparent visual magnitudes match the observed photometry for a range of initial conditions.
The conventional, anti-solar tail observed at 3 au may be present at 4 au but suppressed by 2.6 magnitude in surface brightness by a combination of a decreased production rate and phase angle.
The ice coma of 3I/ATLAS at 4 au resembles a hyperactive coma but
with different rates of sublimation and Haser length scales.
}

 
\maketitle

 \section{Introduction}
   
 A previous analysis \citep{Keto_2025} (KL) of Hubble Space Telescope (HST) images of the interstellar comet 3I/ATLAS   
 at 4 au \citep{Jewitt_2025} showed surface brightness profiles with an exponential dependence
 on projected radius  in both the sunward and perpendicular directions. This earlier study interprets the exponential profiles
 in terms of a Haser model modified for the destruction of icy scattering particles by sublimation. 
 Here we extend our previous results to study the evolution of the ice coma as the comet approaches the Sun,
 and the Haser length scale decreases with the increasing rate of sublimation.

We estimate
the relative contributions to the total effective optical scattering cross sections from the volatile ice and refractory dust 
components of a single population of dirty ice grains.
We describe the grain albedo as the sum of a baseline refractory albedo and an excess due to the higher
optical scattering efficiency of ice. We modify the Haser model to apply to the exponential decay
of the grain albedo from its initial value to its refractory value as the ice is destroyed by sublimation.

When the comet is at far heliocentric distances, the ice with its higher albedo contributes more to the scattered flux from the coma.
Within $\sim 3.5$ au, more rapid sublimation shrinks the ice survival (Haser) length scale, leaving the grains
with only their
refractory component throughout most of the coma.
We calculate the total scattering cross-section of the coma, accounting for the spatial
distribution of the grain albedo and compare the modeled and observed apparent visual magnitudes.

We discuss the results of the calculations and their dependence on the model parameters. We discuss
further aspects of the coma evolution: a conventional tail in the
anti-solar direction, not apparent at 4 au,  was detected around 3 au 
\citep{Jewitt_Luu_2025,Hoogendam_2025}, and the coma 
transitioned to a hyperactive state near perihelion \cite{Tan_2026}.

\subsection{Single-population scattering cross-section with evolving albedo}

We consider a single population of dirty ice grains with a mixed composition of volatile ice and refractory dust
whose phase-free albedo evolves with time as the ice 
sublimates from the grain surface. If $\omega_0$ is the initial albedo of the dirty ice mixture at launch, 
and $\omega_{\rm dust}$
is the albedo of the pure dust component, then the albedo of a grain of size $a$ in a constant velocity
outflow evolves with radial distance $r$ from the nucleus  as,
\begin{equation}
\omega(r,a) = \omega_{\rm dust}
+ \left(\omega_0 - \omega_{\rm dust}\right)
\exp\!\left[-\frac{r}{L(a)}\right].
\end{equation}
The factor $\omega_0 - \omega_{\rm dust}$
represents the increased albedo due to the higher optical scattering efficiency of the ice above the lower baseline albedo of the dust.
The characteristic length, $L(a)$, for the exponential decline is defined by the sublimation time scale, $\tau_{\rm sub}(a)$,
and ouflow velocity, $v_\infty(a)$,
\begin{equation}
L(a) = v_\infty(a)\,\tau_{\rm sub}(a).
\end{equation}
 
The differential scattering cross section contributed by grains of size $a$ inside 
a spherical aperture of radius $\rho_{\rm ap}$ is
\begin{equation}
dC_\omega(a)
=
\int_{V(\rho_{\rm ap})}
n(r,a)\,
\sigma_{\rm sca}(a)\,
\omega(r,a)\, dV .
\end{equation}

The spherical geometric factors in the integral cancel with those in the conservation
equation, leaving $\omega(r,a)$ as
the integrand, 
\\ \\
\( \displaystyle
\int_0^{\rho_{\rm ap}} \omega(r,a)\,dr =
\)
\begin{equation}
\omega_{\rm dust}\,\rho_{\rm ap}
+ \left(\omega_0 - \omega_{\rm dust}\right)
L(a)\left[1 - e^{-\rho_{\rm ap}/L(a)}\right].
\end{equation}

The differential contribution to the total effective optical scattering cross-section from grains of size $a$ is then
\\ \\
\(
dC_\omega(a)
= \displaystyle
\left(\frac{d\dot N}{da}\right)
\sigma_{\rm sca} (a)
\)
\(
\bigg[
\omega_{\rm dust}\, t_{\rm cross}(a)
+ 
\)
\begin{equation}
\left(\omega_0 - \omega_{\rm dust}\right)
\tau_{\rm sub}(a)
\left(1 - e^{-t_{\rm cross}(a)/\tau_{\rm sub}(a)}\right)
\bigg].
\end{equation}
where $d\dot N/da$ is the grain production rate per unit size, and the crossing time is
$t_{\rm cross}(a) = {\rho_{\rm ap}} / {v_\infty(a)}$.

Integration over the grain size distribution, $ \propto a^{-k}$, 
yields the total effective optical scattering cross section inside the aperture,
\\ \\
\(
C_\omega(r_h;\rho_{\rm ap})
=
\) \\ \\
\( \displaystyle
\int_{a_{\min}}^{a_{\max}}
\left(\frac{d\dot N(r_h)}{da}\right)
\sigma_{\rm sca} (a)
\)
\(
\bigg[
\omega_{\rm dust}\, t_{\rm cross}(a,r_h)
+
\)
\begin{equation}
\left(\omega_0 - \omega_{\rm dust}\right)
\tau_{\rm sub}(a,r_H)
\left(1 - e^{-t_{\rm cross}(a,r_h)/\tau_{\rm sub}(a,r_H)}\right)
\bigg] da ,
\end{equation}
where we indicate the dependencies on heliocentric distance, $r_h$.
We assume $k=4$ for equal mass contribution per decade and minimum and maximum
grain sizes of $a_\mathrm{min} = 10^{-7}$ and $a_\mathrm{max} = 10^{-3}$ m \citep{Rinaldi_2017}.
In contrast to the usual total scattering cross-section, this total effective optical scattering cross-section is necessarily
weighted by the albedo factors for the ice and dust components. 

This scattering cross-section has the usual form of a production rate, $\dot N$, times a residence time.  
The first and second terms in the square brackets are the residence times of the refractory and ice components
weighted by their respective albedo factors. 
Separating the right-hand
side into two integrals with the albedo factors outside of the integration, the formulation is mathematically equivalent to two separate
populations of identical grains each with its own residence time and weighted by its own albedo factor. 

To complete the derivation, the scattering cross-section per grain, $\sigma_{\rm sca}$, 
is defined by the scattering efficiency, $Q_{\rm sca}$,
\begin{equation}
\sigma_{\rm sca}  (a) = \pi a^2 Q_{\rm sca}(a) .
\end{equation}
The scattering efficiency scales as $(2\pi a/\lambda)^4$ in the Rayleigh regime and is
approximately constant in the limit of geometric optics,
\begin{equation}
Q_{\rm sca}(a) \approx \min [ (2\pi a/\lambda)^4, 1 ].
\end{equation}
The optical band is approximated by a single wavelength, $\lambda = 0.5\ \mu\rm m$.

The sublimation timescale of the ice component is defined as
\begin{equation}
\tau_{\rm sub}(a,r_h)
\equiv
\frac{f_1 m_g(a)}{f_2 \dot m_{\rm sub}(a,r_h)}
=
\frac{f_1}{f_2}\,
\frac{\rho_g a}{3\,J_{\rm H_2O}(r_h)} .
\end{equation}
Here,  $m_g$ is the grain mass, $\dot m_{\rm sub}$ is the mass loss rate, $f_1$ is the mass fraction of ice in the grain, $f_2$ is the fractional surface area of ice, $\rho_g$ is the grain density,
and $J_{\rm H_2O}(r_h)$ is the sublimation mass flux evaluated at the equilibrium grain temperature.
We absorb the uncertainty of the ratio $f_1/f_2$ into an effective density,
\begin{equation}
\rho_{\rm eff} \equiv \frac{f_1}{f_2}\,\rho_g ,
\end{equation}
so that
\begin{equation}
\tau_{\rm sub}(a,r_h)
=
\frac{\rho_{\rm eff}\,a}{3\,J_{\rm H_2O}(r_h)} .
\end{equation}
The parameter $\rho_{\rm eff}$ and the initial albedo $\omega_0$ encapsulate the unknown grain microphysics
such as the surface exposure, internal structure, and mixing of ice and refractory material.

The total scattered flux is then written as the sum of the refractory baseline and the
volatile-ice excess contribution,
\begin{equation}  \label{total_flux}
\frac{F}{F_\odot}
=
\frac{\Phi(\alpha)}{\pi}\,
\frac{1}{\Delta^2 r_h^2}
\bigg[ C_\omega(r_H;\rho_{\rm ap})
\bigg] ,
\end{equation}
where $\Delta$ and $r_h$ are the geocentric and heliocentric distances.
The phase function factor $\Phi(\alpha)$  is defined by the
standard empirical phase-darkening law for the change in the visual magnitude, $\Delta m = \beta \alpha$,
as a function of phase angle $\alpha$, and $\beta \sim 0.04$.

\section{Comparison with Observations}

As a function of heliocentric distance, the competition between the rising grain production rate and the exponentially
decreasing survival time of the ice produces a local maximum in the apparent magnitude due to ice
at a heliocentric distance $r_h \sim 3-4$ au.
Inside $\sim 3$ au, more rapid sublimation collapses the
Haser survival time and length scales, and the ice component
contributes little to the scattering.  The total apparent magnitude then
asymptotically approaches the refractory baseline contribution. 

In a plot of magnitude versus heliocentric distance, the production rate and the aperture residence time
move the curves of apparent magnitude in orthogonal
directions.
Changes in the production rate   result
in a vertical shift in the apparent magnitudes. 
Model parameters
that change the sublimation rate result in a horizontal shift of the local peak in the 
apparent magnitude due to ice. 

For illustration, we consider three models with different combinations of grain albedo and ice composition: \\
\indent 1) $\omega_0= 0.08$ and ice = 100\% H$_2$O \\
\indent 2) $\omega_0=0.16$ and ice = 100\%H$_2$O \\
\indent 3) $\omega_0=0.16$ and ice = 99.98\% H$_2$O and 0.02\% CO$_2$

\begin{figure*}[p]
\centering

\begin{minipage}{\textwidth}
  \centering
  \includegraphics[width=0.5\textwidth,trim={0 0 0 0},clip]{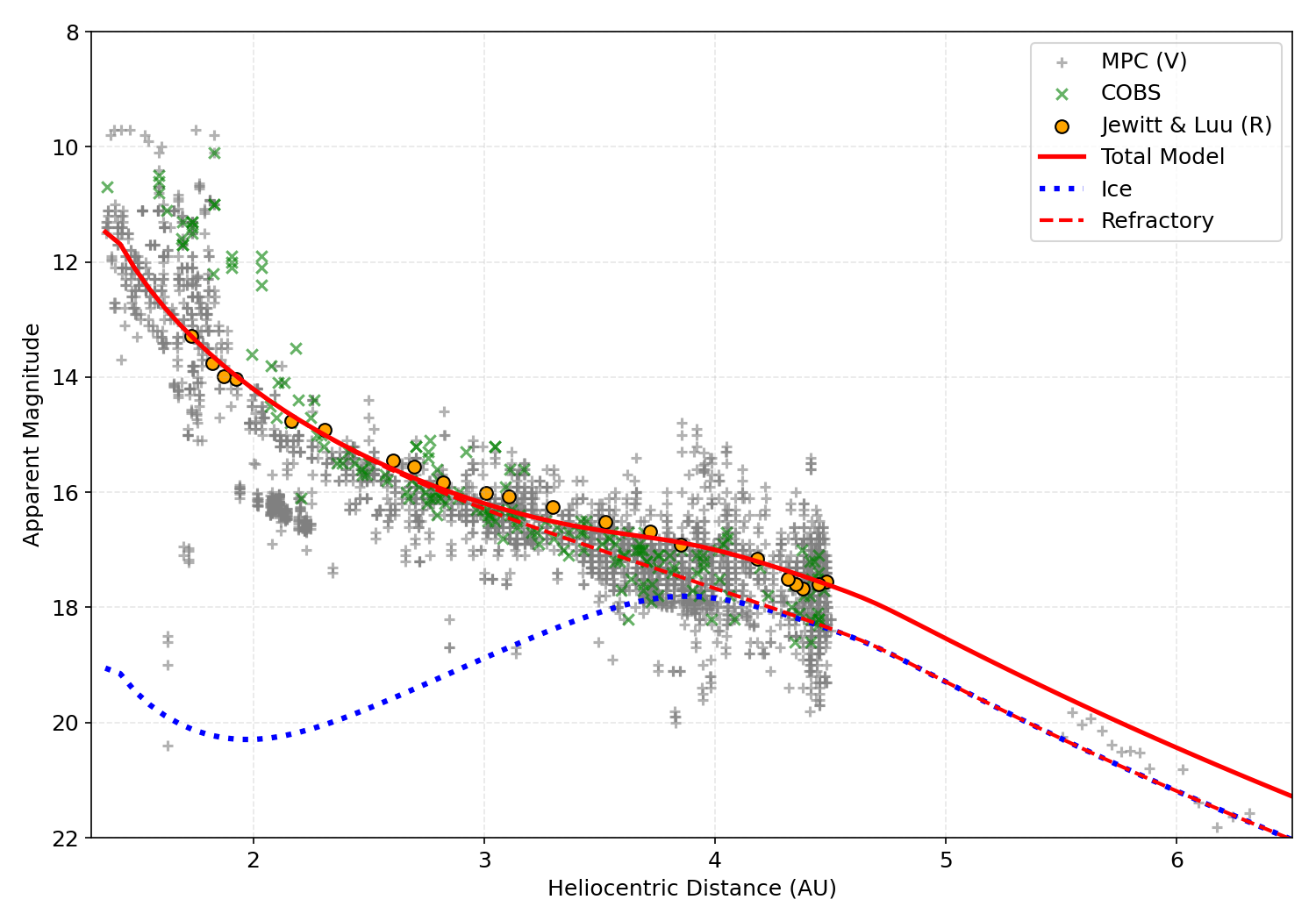}
  \par\smallskip
  \textbf{(a)} $\omega_0=0.08$ and pure H$_2$O ice.
  \label{X}
\end{minipage}

\vspace{0.7em}

\begin{minipage}{\textwidth}
  \centering
  \includegraphics[width=0.5\textwidth,trim={0 0 0 0},clip]{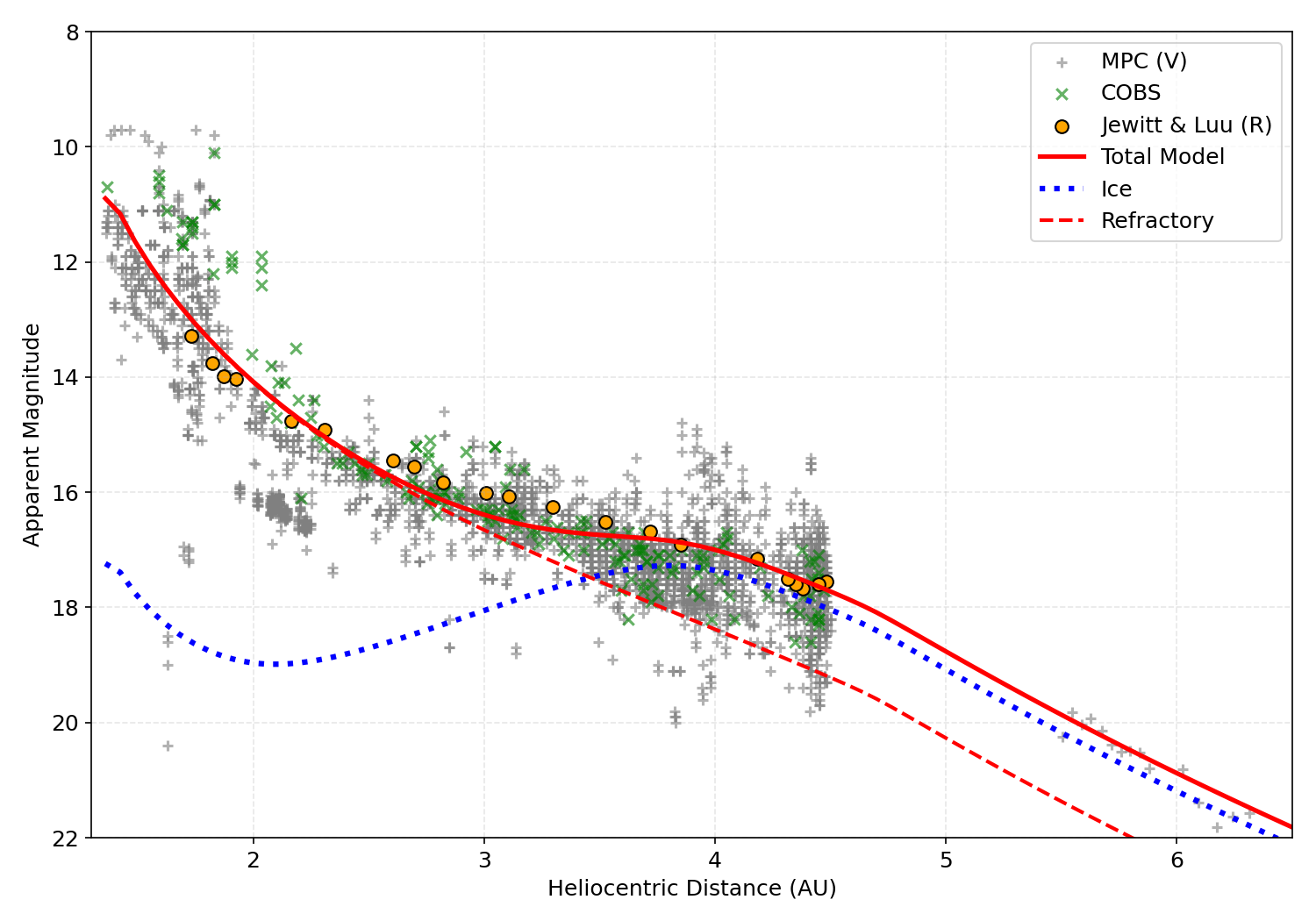}
  \par\smallskip
  \textbf{(b)} $\omega_0=0.16$ and pure H$_2$O ice.
  \label{Y}
\end{minipage}

\vspace{0.7em}

\begin{minipage}{\textwidth}
  \centering
  \includegraphics[width=0.5\textwidth,trim={0 0 0 0},clip]{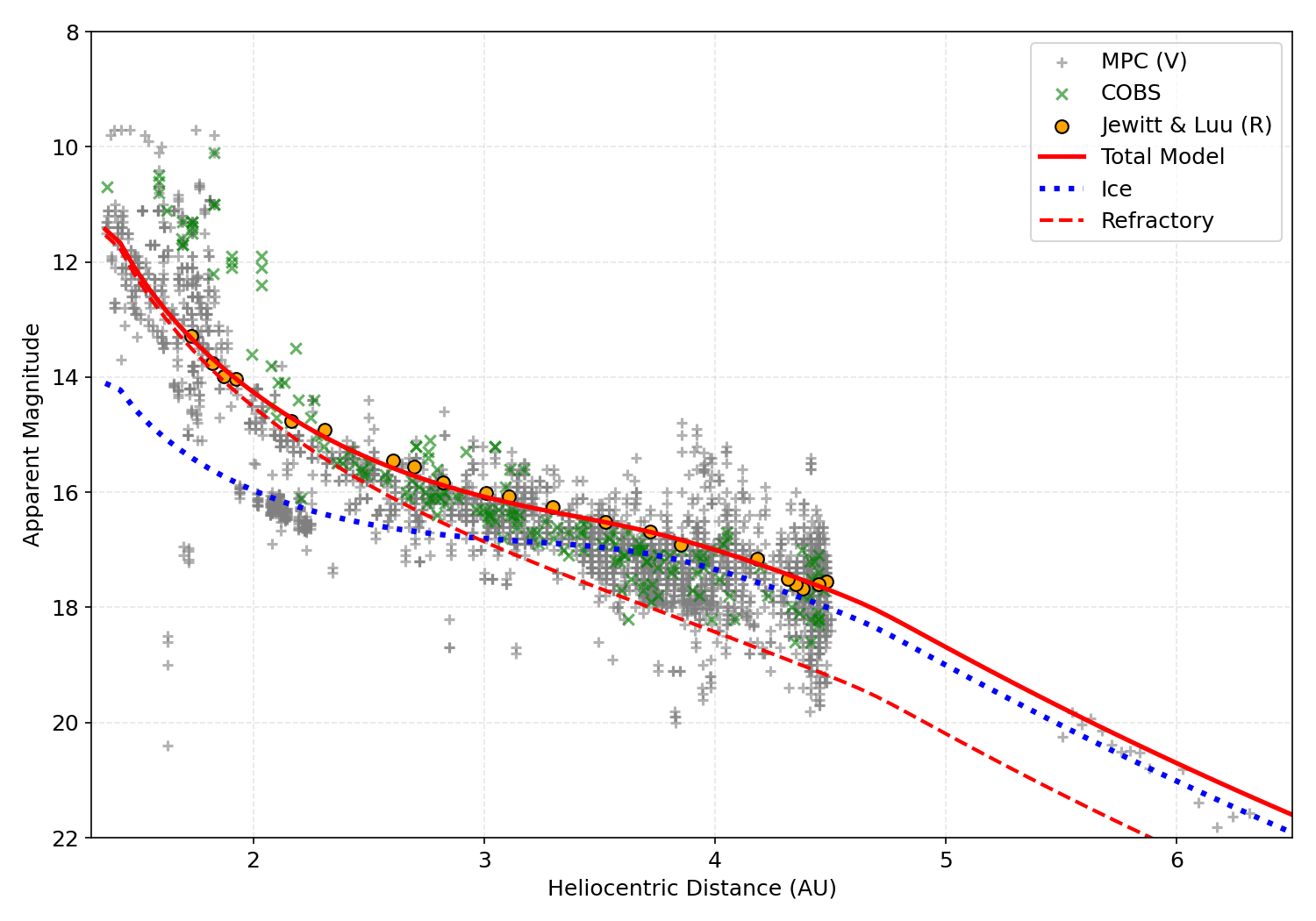}
  \par\smallskip
  \textbf{(c)} $\omega_0=0.16$ and mixed composition ice
  (99.98\% H$_2$O, 0.02\% CO$_2$).
  \label{Z}
\end{minipage}

\caption{Modeled apparent magnitude versus heliocentric distance for three cases. 
Panels (a)--(c) vary the dust baseline albedo and the ice sublimation time scale.}
\label{fig:mag_models}
\end{figure*}

Figures~\ref{fig:mag_models}a -- \ref{fig:mag_models}c compare the apparent visual magnitudes as a function of heliocentric
distance with a compilation of ground-based observations.  \\
\indent (1) Gray crosses show $V$-band apparent magnitudes reported to the Minor Planet Center (MPC),
These include measurements from different telescopes, apertures, and
procedures. \\
\indent (2) Green crosses show photometry from the Comet Observation Database (COBS) subject to the
same heterogeneities as the MPC data. The plotted COBS data are limited to observations with
apertures less than 1 arcminute.\\
\indent (3) Orange circles show $R$-band photometry corrected for
geocentric distance, phase effects, and aperture \citep{Jewitt_Luu_2025}. The plotted data are converted from R to V by  0.45 magnitude. \\
The dashed red curve and blue curves show the refractory and the ice contributions,
respectively.
The total apparent visual magnitude
is shown as a solid red curve.

To set the absolute scaling, we parameterize the unknown grain production rate as a power law,
\begin{equation}
\dot N(r_h) = \dot N_4
\left(\frac{r_h}{4~\mathrm{au}}\right)^{-q},
\end{equation}
where  the grain number production rate at 4 au, $\dot N_4$, is normalized to the precision photometry.  
We use an aperture of constant size $10^4$ km to match the data from \citet{Jewitt_Luu_2025}.
The exponent $q$  
is chosen to best match the overall slope. For models, a, b, c, below, q = 1.5, 2.7, 1.6.

The first two models (figures \ref{fig:mag_models}a and \ref{fig:mag_models}b) show the effect of different values of the initial grain albedo, $\omega_0$.
The range of possible values is from $\omega_\mathrm{dust}$ to about 0.3.  At the lower limit, the ice contributes
nothing to  the grain albedo (buried ice). Some increase due to ice is required to
explain the observed Haser surface brightness profiles. The upper limit is the albedo of pure H$_2$O ice
(ice coated grains) subject
to uncertainties from the structure of the ice itself.  

If the sublimation loss of the higher albedo
ice becomes too rapid at closer heliocentric distances, 
the total apparent magnitude does not increase fast enough to match the observations
at distances around 3 au. The deficit is more apparent in the second model
with an initial $\omega_0 = 0.16$ that results in a total magnitude more dependent
on the ice contribution

The third model shown in figure \ref{fig:mag_models}c explains how to remedy this deficit. This model changes the composition of the
ice from pure H$_2$O to include 0.02\% CO$_2$. The increased cooling from
the latent heat of sublimation of the more volatile CO$_2$ ice lowers the grain temperature
and reduces the rate of sublimation of the H$_2$O ice.
This shifts the local peak of the 
ice scattering to closer heliocentric distances and reduces
the peak sharpness enough to erase this deficit. 

The assumption of a pure ice composition
may be justified by assuming that the more volatile CO$_2$ ice is rapidly destroyed
by sublimation within $\sim1000$ km of the nucleus, consistent with its
sublimation rate at the equilibrium grain temperature. Allowing some residual
CO$_2$ ice is a softening of this assumption potentially explainable by the unknown
structure of the mixture of volatiles and refractories within the grain.

The third example
is representative of the effect of a slower rate of ice sublimation that can
be achieved  through various parameters.
An increase in the maximum grain size or a decrease in the
exponent of the grain size power-law distribution would also result in a shift toward
grains of higher mass with longer sublimation times. 

\section{Discussion}
\subsection{Initial Conditions}

This study and our earlier study (KL) explore whether
the evolution of the coma of 3I/ATLAS can be understood as a consequence of the
physics of comets and their outflows starting  from initial 
conditions required by observations.
The unusual features of this comet,
namely the Haser profiles in scattered light and the prominent anti-tail
suggest
an initial state  with the total effective optical scattering
cross section dominated by volatile ice with an albedo above that of
refractory dust. 

The observations at 4 au and the observed subsequent evolution
are not easily explainable by a coma with
purely refractory grains. Some of the observations are explainable
individually and separately.
For example, the precision photometry
indicates that the apparent magnitudes follow a single power-law 
slope with a production rate scaling as
$r_h^{-3.8}$ \citep{Jewitt_Luu_2025}. As suggested in the reference, this can be 
fit with non-volatile grains, assigning
the increasing production rate to the surface of the nucleus. 

However, the Haser-type surface brightness profiles in the sunward and perpendicular directions
cannot be explained by a coma of refractory grains which to first order
should show profiles scaling with distance from the nucleus as $r^{-1}$.
The exponential profiles observed in different directions
require the destruction of the scattering particles or their optical scattering efficiency
by a process that operates isotropically. 
 
 The observed profiles cannot be successfully reproduced by modification of
an outflow model to include radiation pressure.
The non-constant outflow velocity due to radiative acceleration $a_\mathrm{rp}$ is
$v(r,\theta) = \sqrt{v_\infty^2 - 2 a_\mathrm{rp} r \mathrm{cos} \theta}$
where $\theta$ is the angle from the sunward vector.
The directional deceleration results in density profiles with
direction dependent scaling $\propto 1/(r^2 v(r,\theta))$.
Perpendicular to the sunward direction ($\theta \approx 90$),
the spherical outflow is unaffected.
In the sunward direction, the number density 
increases at the edge of the velocity-limited coma
where  $v(r,\theta) \rightarrow 0$.
The divergence is of course softened by more complete hydrodynamics. Nevertheless,
the observed exponential decline in the surface brightness profiles 
requires a decrease in the number density or the optical scattering efficiency
in all outflow directions.

Furthermore, the characteristic length scale for the radiative deceleration, 
$
r_\mathrm{rp} =  { v_\infty^2 } /  { ( 2 \beta a_\odot /r_h^2  ) }  = 2\times 10^5~(\mathrm{km}),
$
\ is an order of magnitude larger than the Haser length scale (KL).
In this estimate,
the representative 
 grain size is set equal to 
the median scattering-weighted grain size of the distribution, $\sim10$ $\mu$m.
This implies an
estimated terminal velocity of 85 km s$^{-1}$ (KL).

We suggest the following combination of initial properties (KL). \\
  \indent 1. The scattering particles must be volatile ice with a lifetime long enough to travel the extent of the observed anti-tail, 
  	but short enough that the optical scattering efficiency of the ice component of the dirty ice grains
	is diminished by sublimation before solar radiation pressure sweeps the grains into a conventional 
  	tail on the opposite side of the comet. \\
 \indent 2. The sublimation mass flux of gas off the surface of the comet must be rapid enough to maintain the coma, 
 	while the sublimation of the ice grains must be slow enough to meet the lifetime requirement of the first condition. 
 	The necessary ordering of these rates can be achieved if the ice grains are entrained in the outflow 
 	by the gas drag of a more volatile component. \\
 \indent 3. The comet must be rotating slowly enough that the surface temperature responds quickly to the changing
         illumination angle.
 	The mass flux from the sunward surface should continuously exceed that in other directions by a factor sufficient 
 	to produce the observed sunward elongation.
 
This initial state is motivated by several observations. \\
\indent 1. The Haser surface brightness profiles, both in the sunward and perpendicular directions in the HST image, 
	indicate an exponential destruction, a sink,  of the scattering particles or a reduction in their optical
	scattering efficiency with projected distance from the nucleus (KL).  
	This requires the first condition  in the list above. \\
\indent 2. The mixing ratio by number of CO$_2$ to H$_2$O molecules in the gas phase of the coma is 7.6  
         \citep{Cordiner_2025}. Furthermore, the CO$_2$ gas 
	originates from the nucleus while the H$_2$O gas originates from the coma.
	 This motivates the second condition above. \\
\indent 3. The measured rotation period of 3I/ATLAS is $\sim 16.2$ hrs \citep{FuenteMarcos_2025, SantanaRos_2025} while the
	maximum thermal relaxation time scale is 3.7 hrs (KL). This motivates the third condition above.

\subsection{The evolution toward the growth of a conventional, anti-solar tail}

The conventional, anti-solar tail detected for the first time around 3 au was
faint, $\lesssim 10$\% of the coma surface brightness, and short, $\sim$ twice
the coma width. The tail may have existed below the detection limit at 4 au.
Effects that may be responsible for the difference in the brightness of the tail
at 3 and 4 au
include
the grain evolution from icy to refractory, the increase in the production
rates with decreasing heliocentric distance, and the geometric effects of projection
along the orbit.

\paragraph{\bf Grain evolution during sublimation.}
We consider grains initially composed of a mixture of ice and refractory material. 
As the ice sublimates and the grain
mass $m(t)$ decreases, two outcomes
bracket the evolution of the grain structure. Either the grain maintains its initial size, $a_0$ and 
becomes fluffy, or the grain fragments to smaller grains with higher
density. 
We can parameterize the
structural evolution as,
\begin{equation}
a(t) = a_0 \left(\frac{m(t)}{m_0}\right)^{\gamma},
\qquad
\rho(t) = \rho_0 \left(\frac{m(t)}{m_0}\right)^{1-3\gamma},
\end{equation}
where $0 \le \gamma \le \tfrac13 $. 
These scalings affect the
radiation coupling factor,
\begin{equation} \label{beta}
\beta(t) = \frac { Q_\mathrm{rp} } { \rho(t) a(t) } \sim \beta_0\left(\frac{m(t)}{m_0}\right)^{1-2\gamma} .
\end{equation}
The radiation-pressure efficiency $Q_\mathrm{rp}$
varies weakly with the grain composition with slightly higher values for lower grain albedo.
Therefore, for any $\gamma \le \tfrac13$, the effectiveness of the radiative acceleration
parameterized as
$\beta$ increases monotonically with the ice loss. 
If the slope of the grain size distribution is maintained by other fragmentation
processes, the grain evolution may simply shift the distribution to smaller grain sizes.
The grain evolution increases the average $\beta$ of the grains either by means of fluffier 
grains or actually smaller grain sizes. 

The effectiveness of the
grain evolution depends on the mass ratio of ice to dust. 
The scaling of $\beta$ is linear or less, 
with an exponent in a range of $\frac13$ to 1. 
Therefore, an equal mass ratio, volatile to refractory,  
results in only an order unity effect. 
The mass ratio could be arbitrarily higher if heterogeneity in the nucleus
allows the production of cleaner ice grains.

\citet{Jewitt_Luu_2025} 
suggests that the 
development of the conventional tail swept by radiation pressure
would be enhanced by an increase in the average $\beta$ 
resulting from smaller grains at 3 au compared to 4 au. Generally, 
grain sizes evolve from smaller
to larger as a comet approaches the Sun,  following 
Whipple's concept of a maximum-liftable grain size \citep{Whipple_1951}.
However, the structural evolution of sublimating
icy grains provides a natural mechanism to increase $\beta$
in the suggested sense, smaller to larger,
either with fluffier or smaller grains.

\paragraph{\bf Production rates, projection,  and geometric dilution}

Two circumstances combine to increase the detectability of 
a conventional tail at 3 au compared to 4 au. 
First, the observed photometry indicates a production rate increasing
as $r_H^{-3.8}$.  
Second, the phase angle of 3I/ATLAS changed from $9^\circ$ at 4 au to $19^\circ$ at 3 au,
creating a more favorable projection angle to detect a tail.

The ratio of the production rates is,
\begin{equation}
\frac { Q_4\mathrm{au} } {  Q_3\mathrm{au} }  \approx \left( \frac43 \right)^{-3.8} \approx 0.34
\end{equation}
This corresponds to a surface brightness penalty at 4 au, $\Delta m \approx2.5 \log_{10}(3) \approx 1.2$ mag.
Second, the ratio of projection angles requires that the actual distance from the nucleus to a point in the tail
is twice as far at 4 au as it is at 3 au for the same observed distance,
\begin{equation}
\frac { x_4\mathrm{au} } {  x_3\mathrm{au} }  \approx \frac { \mathrm{sin}~19^\circ   }  {  \mathrm{sin}~9^\circ  }  \approx 2
\end{equation}
The surface brightness of the tail decreases with distance from the nucleus due to conical dilution in the outflow.
If we suppose that the surface brightness scales as $\Sigma \propto 1/x$ (rather than $1/x^2$ for spherical), then the 
penalty for dilution is $\Delta m \approx2.5 \log_{10}(2) \approx 0.8$ mag.

While a decreased geocentric distance does not itself affect the surface brightness assuming that the
tail is spatially resolved, the decreased
heliocentric distance increases the surface brightness due to the higher flux of sunlight on the tail
by a factor of $(4/3)^2 \approx 1.8$.  This results in an additional
penalty of $\Delta m = 0.6$ mag at 4 au compared to 3 au. The combined quantified surface brightness
penalty from the production rate and geometric effects is then 2.6 magnitudes. 
Therefore, the faint anti-solar
tail observed at 3 au could exist at 4 au and evade detection.
The additional
unquantified enhanced effectiveness of radiation pressure from the grain evolution
could increase this probability.

\subsection{3I/ATLAS as a hyperactive comet}

Water production rates inferred from a number of telescopes and compiled in \citet{Tan_2026}
show a dependence on aperture size.  This implies that the gas phase water derives
from the extended coma rather than the nucleus, for example from sublimation off icy grains.
\citet{Tan_2026} further suggest that the rapid rise in the production rate as the comet 
approached perihelion implies an expanding active surface area consistent with
increasing sublimation off grains in the coma. 
They suggest that 3I/ATLAS at perihelion is a hyperactive comet,
analogous to 103P/Hartley 2. In their scenario, following \citet{Ahearn_2011,Sunshine_2021}, 
H$_2$O ice grains entrained in the sublimation
mass flux from a more volatile ice in the nucleus such as CO$_2$ 
provide the additional area that becomes active as the comet crosses the 
water-ice sublimation line in heliocentric distance.

Our calculated evolution of the ice-coma model for 3I/ATLAS 
describes this scenario in more detail. From our earlier study (KL),
the ice coma in the initial state at 4 au consists of ice-rich
grains entrained in the sublimation outflow of CO$_2$ gas. 
At 4 au, the equilibrium temperature of the surface of the nucleus $\sim 110 $ K is
set by the solar insolation, re-radiation, and the latent heat of sublimation of CO$_2$.
The sublimation mass flux of H$_2$O off the nucleus is orders of magnitude below
that of CO$_2$. 

The Haser length scale or survival time scale inferred from the observations at 4 au
requires a
rate of sublimation of the grains
consistent with an equilibrium temperature set by 
the sublimation cooling of H$_2$O rather than CO$_2$ (KL). In other words, the icy grains
are essentially pure H$_2$O ice. This is possible regardless of the initial 
composition of the ice because the survival time of the CO$_2$ ice is orders of
magnitude less than the H$_2$O ice. The icy grains lose their CO$_2$ ice on
their own Haser length scale comparable to the radius of the nucleus $\sim 1000$ km.

In the initial state at 4 au, the ice coma represents a frozen hyperactive
state in the following sense. All the production of H$_2$O gas derives from sublimation in the coma
off ice entrained in the CO$_2$ outflow. This is similar to an actual hyperactive state 
that also
requires production of H$_2$O gas in the coma. In the hyperactive case, this is in addition to
the production from the nucleus which is negligible in the frozen state. 
The transition from a frozen to an active hyperactive state
occurs at a heliocentric distance where higher equilibrium grain temperatures 
and more rapid sublimation collapse the Haser length scale of the ice grains in the coma, and a higher surface
temperature activates H$_2$O production from the nucleus.

This possibility does not imply
the initial state as a necessary condition. Hyperactive comets may
form without an initial frozen stage if, for example, ice grains
are supplied by a spontaneously formed vent that extends
below a more refractory surface layer \citep{Ahearn_2011}.

 \section{Conclusions}
 
This study follows the evolution
of a model coma of dirty ice grains composed of volatile ice and refractory dust. 
The calculations explore whether the model at 4 au remains consistent with subsequent 
observations as
the comet approaches the Sun. In particular, if the sublimation of the ice component
has the characteristic time scale to reproduce the 
observations at 4 au, then the sublimation time scale must necessarily decrease at closer heliocentric
distances. This results in a diminution of the total effective optical scattering cross-section of the ice component.

We adapt the Haser model for the photodissociation of molecules to the loss through sublimation
of the high optical scattering efficiency of the ice component. 
The comparison of the calculated apparent visual magnitudes 
with observed photometry shows that the model can reproduce the observations
with a combination of plausible model parameters.

\bibliography{comets_bib}

 \end{document}